\documentclass[prl,twocolumn,showpacs,superscriptaddress]{revtex4-1}
\usepackage{graphicx,epsfig,psfrag,amsopn,epstopdf}
\usepackage[caption=false]{subfig}
\usepackage{bm}
\usepackage{amssymb}
\usepackage{color}
\usepackage{amsmath}
\usepackage{amstext}
\usepackage{latexsym}
\usepackage[usenames,dvipsnames]{xcolor}
\usepackage[colorlinks=true,citecolor=blue,linkcolor=RubineRed,urlcolor=purple]{hyperref}
\usepackage{enumerate}
\usepackage{braket}
\usepackage{tikz}
\usepackage{tikz-feynman} 
\tikzfeynmanset{compat=1.1.0}
\begin{document}
\title{Hubbard model at U=$\infty$: Role of single and two-boson fluctuations}
\author{Debanand Sa}
\author{Anirban Dutta}
\affiliation{Department of Physics, Institute of Science, Banaras Hindu University, Varanasi-221005, India}
\date{\today}
\begin{abstract}
{We have developed a semi-analytical framework formulated in the canonical fermion representation to investigate strongly correlated electron systems. We consider the U=$\infty$ Hubbard model and used the equation of motion method to calculate the fermion self-energy which has two parts: single and two-boson exchange processes. The emergent bosons here are self-generated local charge and spin-density fluctuations which become strongly time-dependent due to extreme correlations. The computed boson spectral density is a diffusive damped mode with a long tail. The electron self-energy at $d=\infty$ is computed self-consistently. The corresponding fermionic spectral density displays a pronounced coherence peak at $\omega=0$, while its frequency derivative develops a two-peak structure at finite $\omega$. The resistivity shows a linear temperature dependence over a broad range, crossing over to coherent Fermi-liquid behavior at extremely low temperatures.}
\end{abstract}
\maketitle
The physics of strongly correlated electronic systems are poorly understood due to their unusual properties as compared to the conventional wisdom of weakly interacting many-electron systems described by the Landau Fermi liquid theory. In particular, the metallic state in a such system, i.e. high temperature superconductors shows linear resistivity over a wide range of temperature window. It shows no sign of resistivity saturation, continues to rise linearly even beyond the Mott-Ioffe-Regel(MIR) quantum limit. This behavior signals the emergence of an unconventional metallic phase, often referred to as a strange metal\cite{review_sm}, in which the concept of well-defined quasiparticles appears to break down. While conventional theoretical approaches based on long-lived quasiparticles have been remarkably successful in describing weakly correlated systems, they fail to provide a qualitative explanation for these unusual behavior observed in these systems. These observations motivate the search for an alternative theoretical paradigm beyond Landau’s framework, one capable of capturing the essential physics of strongly correlated electron systems.\\
The lattice model which captures the essentials of local electronic correlations is the one-band Hubbard model\cite{hubbard} having the inter-site hopping amplitude $t_{ij}$ and the local Coulomb correlation energy $U$. There are excellent reviews\cite{review_hub,tvr_rev_expt} in the Hubbard model which are useful to understand the strongly correlated electron system. The electronic structure calculations as well as the spectroscopic evidences in cuprate superconductors\cite{sheshadri_24} suggest that $t\approx 0.4$ eV and $U\approx4$ eV. This makes $U/t >> 1$ and there by $U/t\rightarrow\infty$ could be a natural starting point to see the effect of strong local electronic correlations in the system. There is an extensive body of theoretical work has addressed the strong-correlation problem using a variety of approaches, including the Gutzwiller projection technique\cite{gutzwiller}, Hubbard X-operator method\cite{hubbardx}, many auxiliary quantum many-body theories\cite{auxiliary}, DMFT\cite{antoine_rmp}, Schwinger source method\cite{shastri_13}, equation of motion(EOM) method\cite{zubarev,kuzemsky,plakida}, SYK-model\cite{syk}, Planckian metals\cite{planckian}, among others. Due to strong correlations, states with double occupancy are excluded, the involved quantum fields do not become canonical Fermi or Bose fields, rather they have additional local constraints which were in general applied globally. In this limit, using Gutzwiller projected wavefunction, P. W. Anderson proposed\cite{anderson_pfl} projected Fermi liquid which shows Fermi edge like singularity in the system. Among subsequent developments, the following set of work\cite{shastri_17,hasan_24} are worth mentioning. The $U=\infty$ Hubbard model was solved and a $\rho-T$ phase diagram was obtained\cite{shastri_17} which included four different regimes: low-temperature Fermi liquid, two strange metallic phases, and high-temperature classical metallic phase. Fermi liquid has $\rho\sim T^2$ and all other regimes have $\rho\sim T$ with different slopes and intercepts. These regimes are separated by three crossover scales, $T_{FL}$ and other two. More recently, \cite{hasan_24}, an approximate self-consistent theory of the same model has been developed using EOM method and similar results have been obtained. This work has been carried out employing Hubbard-X operator representation. In this representation, the involved both Fermi and Bose quantum fields are inherently non-canonical.\\
Research on strange metals have gained renewed momentum due to the recent observation of two bosonic modes of energies 40 and 70 meV respectively in ARPES measurements\cite{expt_b}. Motivated by recent theoretical\cite{hasan_24} and experimental advances\cite{expt_b}, we develop a self-consistent approach to the $U=\infty$ Hubbard model using the EOM technique, employing canonical Fermi and Bose fields. Note that the method is fully general and can, in principle, be implemented in any spatial dimension. In the present work, however, it is solved self-consistently in infinite dimension, where mean-field theory becomes exact. We compute the single particle canonical fermion Green function which obeys the Dyson equation. The self-energy which involves the local charge and spin fluctuations, has two parts: single and two-boson exchange processes. These correlators are calculated by relating them to current-current correlation function. The resulting set of coupled equations are solved self-consistently and some of the physical quantities such as $\rho-T$ phase diagram is obtained. It is worth mentioning that the two-boson exchange process appears due to the canonical nature of the Fermi and Bose fields whereas it does not appear in the EOM method in the X-operator formalism. To the best of our knowledge, such a process has not been reported previously in the literature.\\
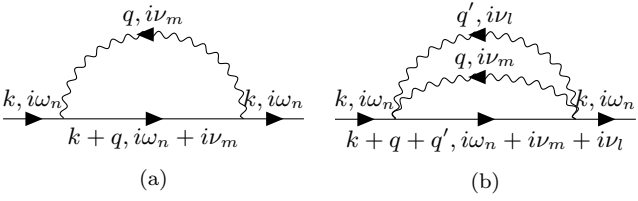
\begin{figure}[t]
\centering
\subfloat[]{\begin{tikzpicture}[
scale=1.,
baseline=(current bounding box.center),
every vertex/.style={inner sep=0pt},
every edge/.style={},
every label/.style={font=\scriptsize},
]
\begin{feynman}
\vertex (a) at (-2.0,0);
\vertex (b) at ( 2.0,0);
\vertex (c) at (-1.2,0);
\vertex (d) at ( 1.2,0);
	
\diagram* {
(a) -- [fermion, edge label=\({k,i\omega_n}\)] (c),
(c) -- [fermion, edge label'=\({k+q,i\omega_n+i\nu_m}\)] (d),
(d) -- [fermion, edge label=\({k,i\omega_n}\)] (b),
(c) -- [anti charged boson, half left, looseness=1.6, edge label=\({q,i\nu_m}\)] (d),
};
\end{feynman}
\end{tikzpicture}}~
\subfloat[]{\begin{tikzpicture}[
scale=1.,
baseline=(current bounding box.center),
every vertex/.style={inner sep=0pt},
every edge/.style={},
every label/.style={font=\scriptsize},
]
\begin{feynman}
\vertex (a) at (-2.0,0);
\vertex (b) at ( 2.0,0);
\vertex (c) at (-1.2,0);
\vertex (d) at ( 1.2,0);

\diagram* {
(a) -- [fermion, edge label=\({k,i\omega_n}\)] (c),
(c) -- [fermion, edge label'=\({k+q+q',i\omega_n+i\nu_m+i\nu_l}\)] (d),
(d) -- [fermion, edge label=\({k,i\omega_n}\)] (b),
(c) -- [anti charged boson, half left, looseness=0.8, edge label=\({q,i\nu_m}\)] (d),
(c) -- [anti charged boson, half left, looseness=1.6, edge label=\({q',i\nu_l}\)] (d),
};
\end{feynman}
\end{tikzpicture}}
\caption{(a) Single and (b) two-boson scattering process.}
\label{fig:b_process}
\end{figure}
\textit{Model Hamiltonian--}We start with the Hamiltonian of the infinite $U$ Hubbard model in its Gutzwiller-projected form as
\begin{eqnarray}
H&=&\sum_{\langle ij \rangle\sigma}t_{ij}(1-n_{i\bar{\sigma}})c_{i\sigma}^{\dagger}c_{j\sigma}(1-n_{j\bar{\sigma}})-\mu\sum_{i\sigma}c_{i\sigma}^{\dagger}c_{i\sigma},
\label{eq:ham}
\end{eqnarray}
where the $c_{i\sigma}^{\dagger}$ is the fermionic creation operator at site $i$ with spin $\sigma$. $t_{ij}$ is the nearest neighbor hopping amplitude and $\mu$ is the chemical potential. We use the Zubarev double-time temperature Green function for fermions, defined as $G_{ij}^{\sigma}(t,t^{\prime})=-i\theta(t-t^{\prime})\langle\left[c_{i\sigma}(t),c_{j\sigma}^{\dagger}(t^{\prime})\right]_{+}\rangle\equiv\langle\langle c_{i\sigma}(t)|c_{j\sigma}^{\dagger}(t^{\prime})\rangle\rangle$. Here the $+$ sign denotes the anticommutator and the expectation value in the grand canonical ensemble. The Heisenberg equation of motion of the above fermion Green function is given by 
\begin{eqnarray}
i\frac{d}{dt}G_{ij}^{\sigma}(t,t^{\prime})&=&\delta(t-t^{\prime})\langle\left[c_{i\sigma}(t),c_{j\sigma}^{\dagger}(t^{\prime})\right]_{+}\rangle\nonumber\\
&&+\langle\langle\left[[c_{i\sigma}(t),H]|c_{j\sigma}^{\dagger}(t^{\prime})\right]\rangle\rangle.
\label{eq:gre1}
\end{eqnarray}
Following the irreducible Green function method developed by Kuzemsky\cite{kuzemsky} and others\cite{plakida} we get,
\begin{eqnarray}
G^{R}_{ij}(\omega)=G_{ij}^{R,MF}(\omega)+\sum_{l}G_{il}^{R,MF}(\omega){\mathcal G}^{R}_{lj}(\omega),
\label{eqn_gr}
\end{eqnarray} 
where $G_{ij}^{R,MF}$ is the mean-field part of the single particle Green function given by $G_{ij}^{R,MF}=\sum_{k} e^{ik\cdot(r_i-r_j)}G_{k}^{R,MF}(\omega)$; $G_{k}^{R,MF}(\omega)=(\omega+\mu-Q\epsilon_{k}+i0^{+})^{-1}$. Here $\epsilon_{k}=-2t\sum_{j}\cos k_j$ is the single particle excitation energy. Within this formulation, the effective single particle bandwidth is renormalized by the factor $Q=1-n/2$, where $n=1-\delta$ is the fermion density and $\delta$ is doping concentration. ${\mathcal G}^{R}_{ij}(\omega)$ is the higher order irreducible Green function originating from the fluctuations due to the commutator of the second term in the RHS of Eq.(\ref{eq:gre1}). The irreducible Green function can be calculated by taking derivative of the same with respect to $t'$ which yields ${\mathcal G}^{R}_{lj}(\omega)=\sum_{l^\prime}{\rm T}_{ll^\prime}^{\sigma}(\omega)G_{l^\prime j}^{\sigma}(\omega)$. Here, ${\rm T}_{ll'}^{\sigma}(\omega)$ denotes the scattering matrix. Upon substituting ${\mathcal G}^{R}_{lj}(\omega)$ back into Eq.~(\ref{eqn_gr}), the full Green function can be written as the sum of the mean-field propagator and the contributions from repeated scattering processes, $G = G^{\mathrm{MF}} + G^{\mathrm{MF}}\, {\rm T}\, G^{\mathrm{MF}}$. The self-energy $\Sigma$ is introduced by identifying it with the irreducible part of the scattering matrix, ${\rm T} = \Sigma + \Sigma\, G^{\mathrm{MF}}\, {\rm T}$.  Consequently, the full Green function satisfies the Dyson equation $G = G^{\mathrm{MF}} + G^{\mathrm{MF}}\, \Sigma\, G$\cite{book}. The self-energy can be written as
\begin{eqnarray}
\Sigma_{ij}^{\sigma}(\omega)&=&\sum_{ll^\prime}t_{il}t_{l^\prime j}\langle\langle(P_{i\bar{\sigma}}P_{l\bar{\sigma}}c_{l\sigma}-c_{i\bar{\sigma}}^{\dagger}c_{l\bar{\sigma}}P_{l{\sigma}}c_{i\sigma})|\nonumber\\
&&(P_{j\bar{\sigma}}P_{l^\prime\bar{\sigma}}c_{l^\prime\sigma}-c_{l^\prime\bar{\sigma}}^{\dagger}c_{j\bar{\sigma}}P_{l^\prime{\sigma}}c_{j\sigma}^{\dagger})\rangle\rangle_{\omega},
\end{eqnarray}
%
where $P_{i\sigma}=1-n_{i\sigma}$ is the projection operator at site $i$ with spin $\sigma$. Since the projection operator is an outcome of strong electronic correlations, this has to be treated quantum mechanically for the case of $U=\infty$, rather than replaced by a static average. In order to simplify the self-energy, we take the spatial dimension $d=\infty$ limit which makes the self-energy local and the mean-field limit becomes exact. Using Wick's expansion, we go for a decoupling approximation for the self-energy which yields two distinct processes at the tree-level (self-generated one-boson and two-boson exchange processes) as shown in the Fig.~\ref{fig:b_process}. This can be understood from the Hamiltonian given by Eq.(\ref{eq:ham}). Rewriting the projected hopping part of the Hamiltonian explicitly gives us four different terms; the first term is the conventional bare hopping term, the second and third ones are the two-body interacting terms where the hopping is coupled to the local electronic densities. And the last term is a three-body term where hopping is coupled to two densities at the neighboring sites. The two-body and the three-body terms are responsible for the above processes shown in Fig.~\ref{fig:b_process}. Thus the self-energy $\Sigma$ can be written as $\Sigma=\Sigma_{1}+\Sigma_{2}$ where,
\begin{eqnarray}
\Sigma_{1}^{R}(\omega)&=&6t^2(1-\frac{3n}{4})\Big[\langle\langle N(t)|N(t')\rangle\rangle_{\omega}\nonumber\\
&&+2\langle\langle S^{+}(t)|S^{-}(t')\rangle\rangle_{\omega}\Big]\langle\langle c(t)|c^{\dagger}(t')\rangle\rangle_{\omega},\\
\Sigma_{2}^{R}(\omega)&=&-2t^2\Big[\langle\langle N(t)|N(t')\rangle\rangle_{\omega} \langle\langle S^{+}(t)|S^{-}(t')\rangle\rangle_{\omega}\Big]\nonumber\\
&&\langle\langle c(t)|c^{\dagger}(t')\rangle\rangle_{\omega}.
\end{eqnarray}
The self-energies $\Sigma_{1}$ and $\Sigma_{2}$ which contain the bosonic fluctuation are the electron charge density $\langle\langle N|N\rangle\rangle$ and spin density $\langle\langle S^{+}|S^{-}\rangle\rangle$ Green functions. $N$ and $S$ are respectively the number and spin-1/2 operators, $N=(n_{\uparrow}+n_{\downarrow})$, $S^{+}=c_{\uparrow}^{\dagger}c_{\downarrow}$, $S^{-}=c_{\downarrow}^{\dagger}c_{\uparrow}$, and $S^{z}=\frac{1}{2}(n_{\uparrow}-n_{\downarrow})$.  In deriving $\Sigma_{2}^{R}(\omega)$, the boson-boson correlations have been further approximated in the mean-field way such that both the bosons are free at the lowest order. The $\langle\langle N|N\rangle\rangle=D_{N}=D_{N}^{+}+D_{N}^{-}$ and $\langle\langle S^{+}|S^{-}\rangle\rangle=D_{S}=D_{S}^{+}+D_{S}^{-}$ Green functions can be computed from current-current correlation functions following the procedure outlined in Hassan et al.\cite{hasan_24}.
Using the spectral representation for the above Green functions the retarded local self-energies can be written as
\vspace{-0.7em}
\begin{widetext}
\begin{eqnarray}
\Sigma_{1}^{R}(\omega)&=&6t^2(1-\frac{3n}{4})\iint d\epsilon_1 d\epsilon_2 \Big[\rho_{D_{N}}(\epsilon_1)+2\rho_{D_{S}}(\epsilon_1)\Big]\rho_G(\epsilon_2)\frac{\left[n_B(\epsilon_1)+n_F(\epsilon_2)\right]}{\omega+\epsilon_1-\epsilon_2+i0^+},\\
\Sigma_{2}^{R}(\omega)&=&-(2t^2\beta)\iiint d\epsilon_1 d\epsilon_2 d\epsilon \rho_{D_{N}}(\epsilon_1)\rho_{D_{S}}(\epsilon_2)\rho_G(\epsilon)
\frac{\left[\left(n_{B}(\epsilon_2)+n_{F}(\epsilon)\right)\left(n_{B}(\epsilon_1)+n_F(\epsilon-\epsilon_2)\right)\right]}{\omega+\epsilon_1+\epsilon_2-\epsilon+i0^+}.
\end{eqnarray}
\end{widetext}
\vspace{-0.7em}
Here $\rho_G(\epsilon_2)=-{\frac{1}{\pi}\mathrm{Im}G^{R}(\epsilon_2)}$, $\rho_{D_{N}}(\epsilon_1)=-{\frac{1}{\pi}\mathrm{Im}D_{N}^{R}(\epsilon_1)}$, $\rho_{D_{S}}(\epsilon_1)=-{\frac{1}{\pi}\mathrm{Im}D_{S}^{R}(\epsilon_1)}$. $n_B$ and $n_F$ are Bose and Fermi distribution functions respectively and $\beta$ is the inverse temperature.\\
The boson Green functions are calculated using the EOM method similar to that of Hassan et al.\cite{hasan_24}
\begin{eqnarray}
D_{N}^{\alpha}(\omega)=\alpha\frac{n\omega+\chi_{\gamma}^{\alpha}}{\omega^2};~~~D_{S}^{\alpha}(\omega)=\alpha\frac{(n/2)\omega+\chi_{\gamma}^{\alpha}}{\omega^2},
\end{eqnarray}
where $\alpha=\pm$ and $\chi_{\gamma}^{\alpha}(\omega)$ are the charge and spin current-current correlation functions whose spectral representation is given by 
\begin{eqnarray}
\chi^{\alpha}_{\gamma}(\omega)&=\int_{-\infty}^{\infty} d\omega^{\prime} \frac{\left[\frac{1+\alpha}{2}+n_B(\omega^\prime)\right]\rho_{\gamma}(\omega^\prime)}{\omega-\omega^\prime+i0^{+}},
\label{eq_chi}
\end{eqnarray}
where $\gamma$ is either $N/S$ and $\rho_{\gamma}(\omega^\prime)=-\frac{1}{\pi}\mathrm{Im}\chi_{\gamma}^{R}(\omega')$. The $\chi^R(\omega)$ is calculated from the particle-hole bubble diagram\cite{book}. Using the relation $\chi_{N}^R(\omega)=2\chi^R(\omega)$ and $\chi_{S}^R(\omega)=\chi^R(\omega)$ the spectral representation for the current-current correlation function is written as
\begin{eqnarray}
\chi^{R}(\omega)&=& \frac{1}{N} \sum_{k} \int d \omega_{1} d \omega_{2} \frac{\rho_{G}(k,\omega_{1}) \rho_{G}(k,\omega_{2}) v_{k}^{2}}{\omega + \omega_{1} - \omega_{2} + i0^+} \nonumber\\
&&\times\Big (n_{F}(\omega_{2}) - n_{F}(\omega_{1})\Big),
\label{eq:chi}
\end{eqnarray}
where $\rho_{G}(k,\omega_{2})=-\frac{1}{\pi}\mathrm{Im}G^R(k,\omega)$. The imaginary part of $\chi^R(\omega)$ can be calculated adopting a Bethe lattice involving transport density of states\cite{hasan_24}. The fermionic Green function, including its self-energy, and the bosonic Green functions were computed self-consistently, as outlined in the Supplementary Material.\\
\begin{figure}[h!]
\subfloat[~]{\includegraphics[width=0.494\columnwidth]{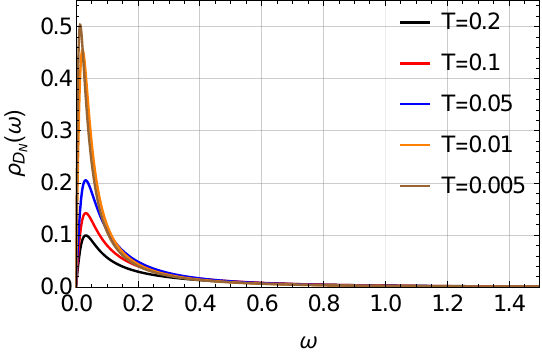}\label{fig:2a}}
\subfloat[~]{\includegraphics[width=0.494\columnwidth]{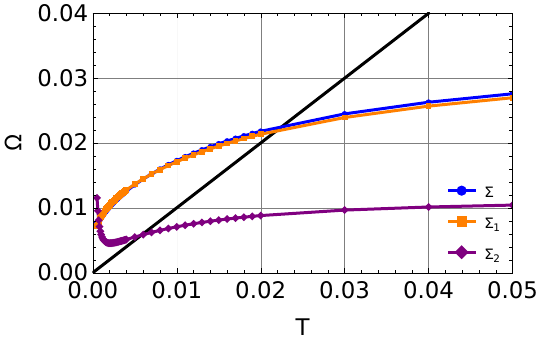}\label{fig:2b}}
\caption{Bosonic fluctuation generated due to strong correlation at $n=0.7$. (a)Bosonic spectral function $\rho_{D_{N}}(\omega)$ as a function of $\omega$ and (b) boson frequency(blue) as a function of temperature. Orange and purple line indicate boson frequency associated with $\Sigma_{1}$ and $\Sigma_{2}$ respectively and black line is $\Omega=T$ straight line. The intersection of $\Omega_{1}$ and $\Omega_{2}$ with the $Y=T$ straight line provides two frequency scales respectively.}
\label{fig:2}
\end{figure}
\textit{Results--}The bosonic charge/spin spectral density functions which arise from the self-generated local charge/spin density fluctuations are inevitable to appear due to the physics of strong correlations. They coupled to the electronic degrees of freedom and the electron dynamics then is the cause for incoherence giving rise to $T$-linear resistivity. The charge spectral density as a function of $\omega$ at different temperatures is shown in Fig.~\ref{fig:2a} which is antisymmetric with $\omega$. This is a massless damped excitation having broad asymmetric peak with a long tail. The reason for such an excitation is the charge at each lattice site diffuses quantum mechanically in a dynamic way. Since there are two distinct boson processes(one boson exchange and two-boson exchange) the typical energy scales in both the spectral densities are calculated as, $\Omega_{i}(T)=\int_{\omega_{l}}^{\omega_{u}} d\omega\omega\left(\frac{\rho_{D_{N_i}(\omega)}}{\omega \kappa}\right)$ where $i=1,2$ and $\kappa$ being the compressibility, $\kappa=\int_{-\infty}^{\infty}d\omega\left(\frac{\rho_{D_{N}(\omega)}}{\omega}\right)$. Here the frequency window is chosen as $\omega_{l}=0.002t$ and $\omega_{u}=30t$. The $T$ dependence of $\Omega_{i}(T)$ is shown in Fig.~\ref{fig:2b}. At low $T$, $\Omega_{1}(T)$ increases sub-linearly with $T$, whereas $\Omega_{2}(T)$ shows upturn and both flatten out at higher $T$. The classical limit for both the fluctuations are determined from an equipartition energy curve $Y=T$, where the frequency scale for two-boson process($\Omega_{2}$) is lower than that of one-boson($\Omega_{1}$) process, indicating that two-boson fluctuations become energetically relevant at comparatively lower frequencies.\\
\begin{figure}[t]
\subfloat[~]{\includegraphics[width=0.494\columnwidth]{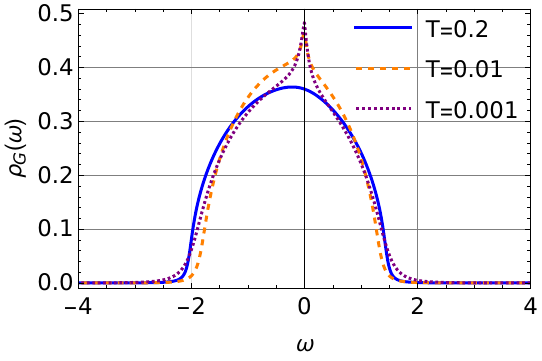}\label{fig:3a}}
\subfloat[~]{\includegraphics[width=0.494\columnwidth]{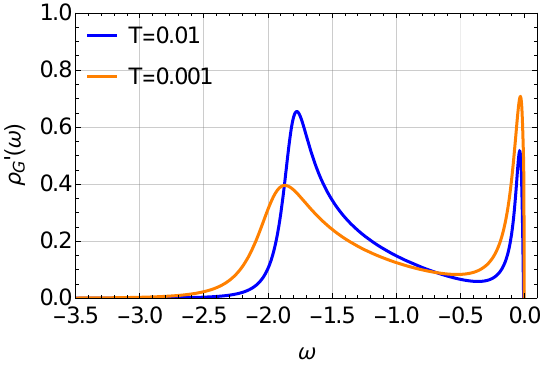}\label{fig:3b}}\\
\subfloat[~]{\includegraphics[width=0.494\columnwidth]{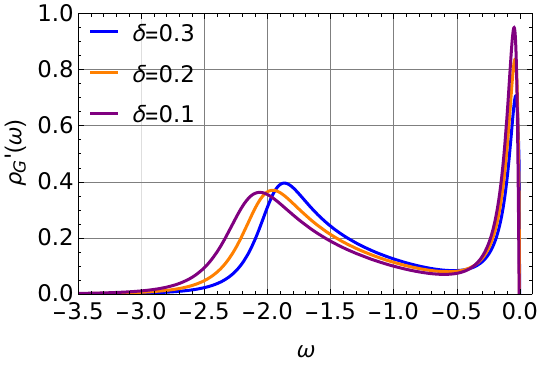}\label{fig:3c}}
\subfloat[~]{\includegraphics[width=0.494\columnwidth]{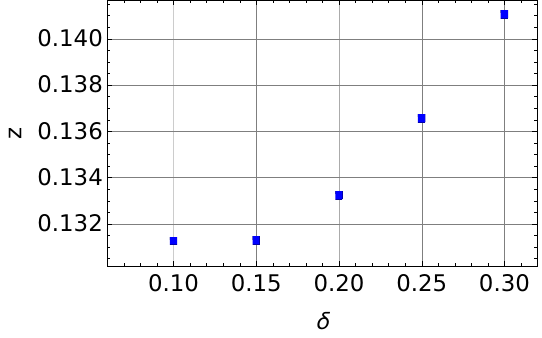}\label{fig:3d}}
\caption{Spectral function of strongly correlated electrons. (a)Spectral function at different temperature, as a function of $\omega$. (b)First derivative of spectral function with $\omega$ at different temperatures and (c) different doping concentrations. (d)The variation of quasiparticle weight with doping.}
\label{fig:3}
\end{figure}
\begin{figure}[b]
\subfloat[~]{\includegraphics[width=0.494\columnwidth]{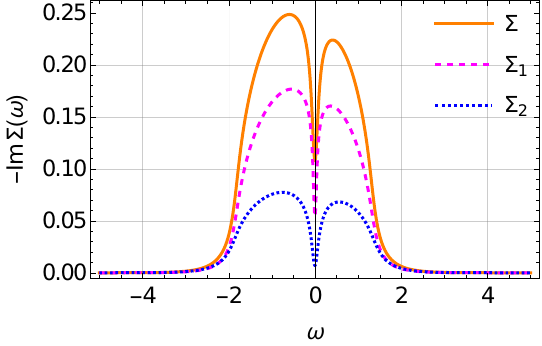}\label{fig:4a}}
\subfloat[~]{\includegraphics[width=0.494\columnwidth]{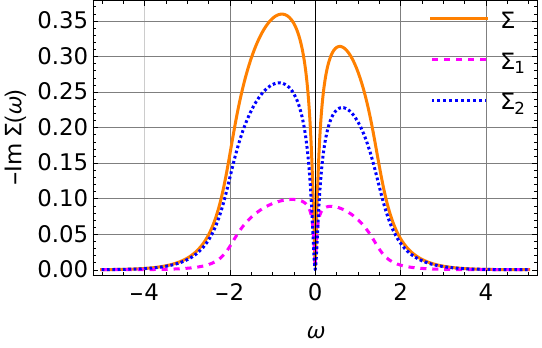}\label{fig:4b}}\\
\subfloat[~]{\includegraphics[width=0.494\columnwidth]{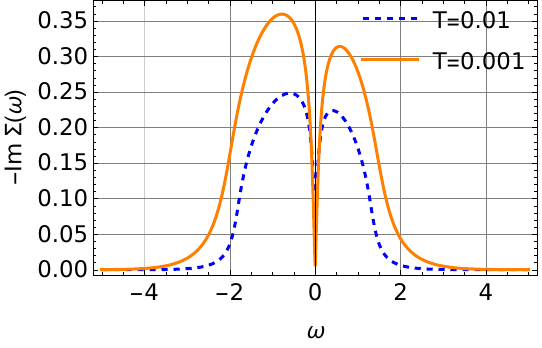}\label{fig:4c}}
\subfloat[~]{\includegraphics[width=0.494\columnwidth]{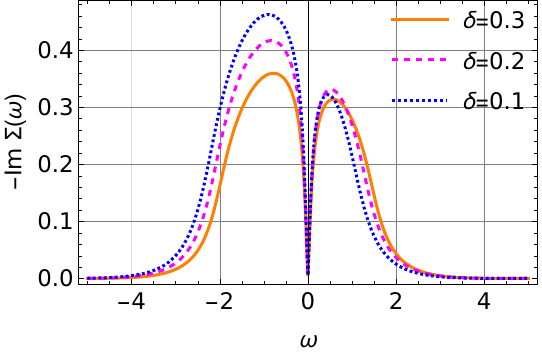}\label{fig:4d}}
\caption{The dip in the self-energy is due to coherence. Imaginary part of the self-energy is plotted with $\omega$ for (a)$T=0.01$ and (b) $T=0.001$. The contributions of the two distinct processes to the self-energy are plotted separately and at low temperature, contribution from $\Sigma_{2}$ is dominant. The total self-energy for different (c)temperatures and (d) doping is plotted with $\omega$.}
\label{fig:4}
\end{figure}
The imaginary part of the electronic self-energy provides insights into the lifetime and the coherence of the quasi-particles which are crucial for the understanding of the low energy physics of the material. Fig.~\ref{fig:3} and \ref{fig:4} summarizes some of the main results.  At extremely low temperatures, a pronounced coherence peak emerges at $\omega=0$ in the fermionic spectral density. This feature corresponds to dips in $\Sigma_1$, $\Sigma_2$, and the total self-energy $\Sigma$, signaling long-lived quasiparticle excitations. As the temperature increases, the coherence peak progressively broadens and eventually disappears, marking a crossover to an incoherent metallic state. The broadening in the spectral densities are due to its coupling to the incoherent bosons which are seen from the two-peak structure in the first derivative $\rho_{G}^{\prime}(\omega)$. Notably, such two-peak structure has been observed in the recent ARPES measurements\cite{expt_b}. The evolution of $\rho_{G}^{\prime}(\omega)$ with frequency at different temperature and doping concentrations are shown in Fig.~\ref{fig:3b} and \ref{fig:3c}. At extremely low temperature, the contribution of $\Sigma_1$ and $\Sigma_{2}$ to the coherence peak is almost similar but, the contribution from $\Sigma_{2}$ dominates over $\Sigma_{1}$ as shown in Fig.~\ref{fig:4b}. However, at higher temperature the contribution from $\Sigma_{1}$ dominates over $\Sigma_{2}$ and both contributes to the coherent and incoherent behavior shown in Fig.~\ref{fig:4a}. The frequency dependence of $\mathrm{Im} \Sigma(\omega)$ at various temperatures and doping concentrations are shown in Fig.~\ref{fig:4c} and \ref{fig:4d}, highlighting the evolution from a sharply defined low-energy structure to a broadened, incoherent spectrum at elevated temperatures.\\
\begin{figure}[b]
\subfloat[~]{\includegraphics[width=0.494\columnwidth]{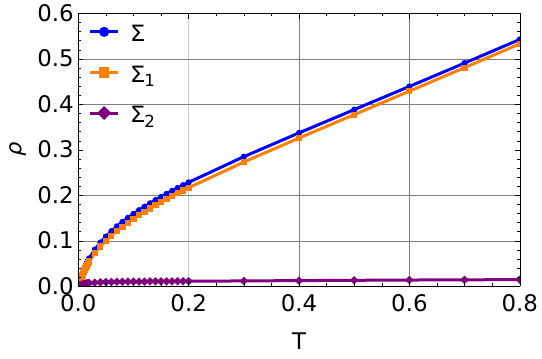}\label{fig:5a}}
\subfloat[~]{\includegraphics[width=0.494\columnwidth]{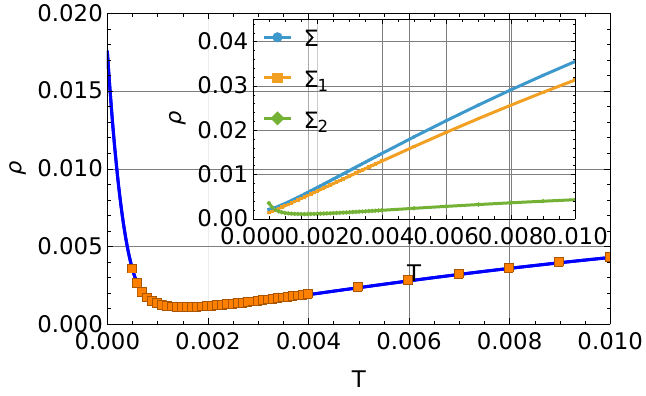}\label{fig:5b}}\\
\subfloat[~]{\includegraphics[width=0.494\columnwidth]{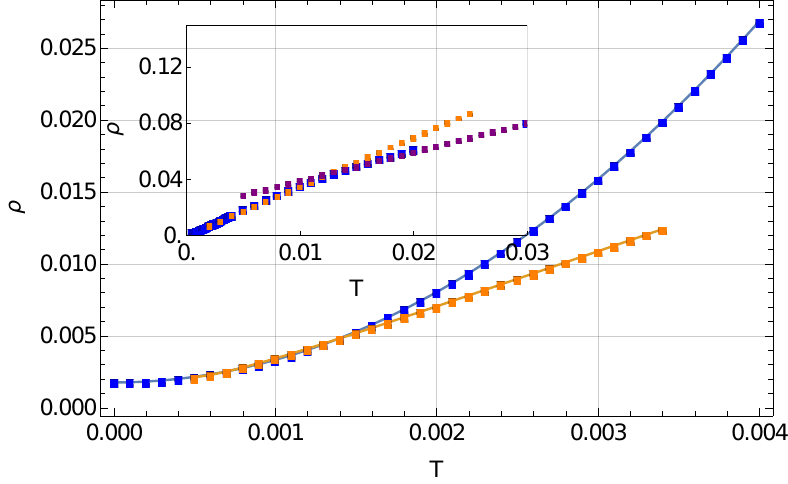}\label{fig:5c}}
\subfloat[~]{\includegraphics[width=0.494\columnwidth]{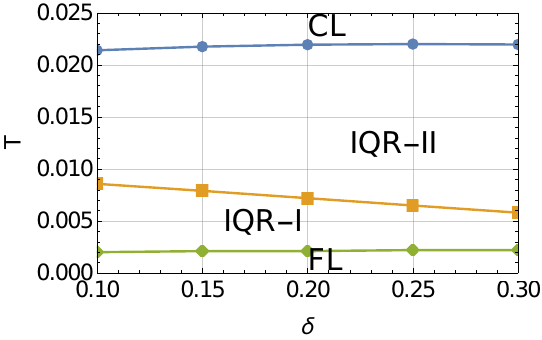}\label{fig:5d}}
\caption{DC resistivity showing linear as well as quadratic $T$ behavior. (a) Resistivity is plotted with temperature showing linear scaling. The contribution from $\Sigma_{2}$ to resistivity show upturn at very low temperature, shown in inset (b). The low temperature behavior is magnified in (b) with blue line indicate spline interpolation of the data. (c) At extreme low temperature, the resistivity data(orange) is fitted with $T^2$(blue) to extract $T_{FL}$. Two linear fits are shown in the inset. (d) The phase-diagram from our numerical simulation.}
\label{fig:5}
\end{figure}
Within our framework, the DC resistivity is computed from the imaginary part of the current-current correlation function $\mathrm{Im}\chi_{N}(\omega)$ and its $T$-dependence is plotted in Fig.~\ref{fig:5a}. It has a large temperature window showing linear behavior and no signs of resistivity saturation beyond the MIR limit. However, at extremely low temperature, it shows Fermi liquid behavior similar to mentioned in these literature\cite{dmft_fl,shastri_17,hasan_24}. We have fitted in Fig.~\ref{fig:5c} the low temperature data with the  Fermi-liquid behavior($T^2$) to extract the Fermi-liquid(FL) scale $T_{FL}$ and the high temperature classical regime($T_{CL}$) is obtained from the frequency scale($\Omega_{1}$) associated with the single-boson process. The intermediate regime in between $T_{CL}$ and $T_{FL}$ is the incoherent quantum regime(IQR). The IQR regime can be fitted with two linear fits with different slopes which provides a new scale $T_{IQR}$. $T_{IQR}$ further splits the IQR regimes into two incoherent regimes, IQR-I and IQR-II, shown in Fig.~\ref{fig:5d}. This looks similar to the proposed $T-\delta$ phase diagram of \cite{shastri_17}. The resistivity has two contributions coming from $\Sigma_{1}$ and $\Sigma_{2}$ as plotted in Fig.~\ref{fig:5}. At extremely low temperature, both $\Sigma_{1}$ and $\Sigma_{2}$ has Fermi-liquid contribution($\propto T^2$) to the resistivity, however the contribution from $\Sigma_{2}$ shows upturn\cite{upturn} as in Fig.~\ref{fig:5b}, indicating the enhanced role of two-boson processes in shaping the low-temperature transport properties.\\
\textit{Discussions--}As emphasized in the introduction, strong correlation makes the Fermi and Bose fields intrinsically non-canonical. Hassan et al.\cite{hasan_24} employed these non-canonical Fermi and Bose fields in their EOM method, whereas Shastry et al.\cite{shastri_17} adopted the Schwinger-source method formalism. Despite methodological differences, both approaches yield qualitatively similar physical conclusions: an extended incoherent regime over a wide temperature range, which eventually crosses over to a coherent Fermi-liquid state at very low temperatures. In contrast, the present work reformulates the EOM approach entirely in terms of canonical fermionic and bosonic fields, in which two distinct boson exchange processes appear in the fermion self-energy. The derivative in the calculated spectral density shows two boson peak structure similar to that observed in the recent ARPES measurements. Moreover, at extremely low-$T$ the two-boson exchange process has prominent Fermi-liquid contribution in addition to $\Sigma_{1}$. Our results may also be contrasted with that of Arnold et al.\cite{kopietz}, who have recently developed a functional renormalization group study for the $U=\infty$ Hubbard model in a 2D square lattice. They proposed a phase diagram for the ground state which crosses from the incoherent non-Fermi liquid state to Fermi-liquid as the electron density varies from large to small. There is a recent work in the area of strongly correlated electron systems which deals with the coupled electron-boson model where the bosons are critical couples to fermion via Yukawa interaction with static disorder\cite{aavishkar}. Some of the broad results obtained by them are the linear resistivity and also the crossing over to the low-$T$ Fermi-liquid regime. While these approaches successfully capture several experimental trends, but our approach is material based strong coupling over-simplified lattice model. Although simplified, it provides a microscopic strong-coupling perspective in which the fluctuation-mediated processes arise self-consistently from the underlying Hubbard dynamics.\\
\textit{Conclusions--} In conclusion, in the present letter, we have investigated the problem of $U=\infty$ Hubbard model at $d=\infty$ limit which unmasks the complex dynamics in the extremely correlated electron systems. Due to the strong local constraints of no double occupancy, electron couples to the local self-generated damped density fluctuations giving rise to a incoherent behavior in the resistivity in a wide temperature range and crosses over to a coherent regime at extremely low temperature. The variation of electron self-energy, spectral-density, resistivity and the bosonic spectral density with frequency, temperature and doping supports the above physical picture. In particular, the interplay between single and two-boson exchange processes plays a central role in shaping both
the incoherent and coherent regimes. However, these results may not directly correlate with the observation from the specific systems due to the fact that the present theory is for $U=\infty$ and $d=\infty$. In finite dimensions, i.e., 2D in the case of high temperature superconductors, one has to consider the $1/d$ corrections in the present formulation. Most of the strongly correlated systems have large but finite $U$ leading to $1/U$ perturbative effects giving rise to a nearest neighbor anti-ferromagnetic super-exchange integral $J(t^2/U)$. This might provide a new low energy scale which is beyond the scope of the present work. We would like to explore along this direction in our future work. Our letter opens up several avenues for further numerical and theoretical investigations. Overall, our study provides a strong-coupling, self-consistent framework for understanding incoherence and strange metallic behavior in $U=\infty$ Hubbard model, and it opens several avenues for further analytical and numerical investigations beyond the infinite-$U$ and infinite-dimensional limits.\\

\textit{Acknowledgements}--We sincerely thank Prof. T. V. Ramakrishnan for numerous insightful and stimulating discussions that were instrumental in developing the ideas presented in this work. We also acknowledge Prof. S. R. Hassan for several invaluable discussions.\\
We thank IoE BHU for financial support. A. Dutta acknowledges financial support from the SERB(ANRF)-SRG Grant No. SRG/2022/001145.
\\
\onecolumngrid
\appendix 
\section{Supplimetary material}
\subsection{Self-consistent numerical implementation}
Here we briefly outline the essential steps and equations required for the numerical implementation of the semi-analytical scheme described in the main text. In the presence of interactions, the momentum-resolved Green function is given by
\begin{eqnarray}
G_{k}^{\sigma }(\omega)=\frac{1}{\omega+\mu-Q\epsilon_k-\Sigma(k,\omega)},
\end{eqnarray}
where $\Sigma(k,\omega)$ is the total self-energy $\Sigma(k,\omega)=\Sigma_1(k,\omega)+\Sigma_2(k,\omega)$. In infinite dimension, self-energy is purely site-local and in this limit, the local Green function can then be computed as,
\begin{eqnarray}
G_{loc}(\omega)=\int \frac{\rho_0(\epsilon)d\epsilon}{\omega+\mu-Q\epsilon-\Sigma},
\label{eq:g_loc}
\end{eqnarray}
where $\rho_0(\epsilon)$ is the semi-circular density of states of the Bethe lattice. 
The fermionic spectral function is obtained from $\rho_{G}(\omega)=-\frac{1}{\pi}\mathrm{Im}{G_{loc}(\omega)}.$ From the expression of $\chi^{R}(\omega)$ given in Eq.(\ref{eq_chi}) of the main text, the imaginary part can be expressed as
\begin{eqnarray}
\mathrm{Im}[\chi^{R}(\omega)]& = - \pi \int d\epsilon  d\omega_{1} \Phi(\epsilon) \rho_{G}(\epsilon,\omega_1) \rho_{G}(\epsilon,\omega + \omega_{1}) \left[n_{F}(\omega_{1}) - n_{F}(\omega + \omega_{1})\right],
\label{curr_corr_exp}
\end{eqnarray}
where $\Phi(\epsilon) = \frac{1}{N} \sum_{k} v_{k}^{2} \delta(\epsilon - \epsilon_{k}) = \Phi_{0} (4-\epsilon^{2})^{3/2}$ and $\rho_{G}(\epsilon,\omega)=-\frac{1}{\pi}\mathrm{Im}[{G(\epsilon,\omega)}]$. The above convolution integral is evaluated using the convolution/correlation theorem. The real part of $\chi$ has been computed from the imaginary part using Kramers-Kronig relations. Using the expression for $\chi^{R}_{N}(\omega)$ and $\chi^{R}_{S}(\omega)$, we can express the imaginary part of $\chi^{\alpha}_{ \gamma}(\omega)$ given in Eq.(\ref{eq_chi}) as
\begin{eqnarray}
\mathrm{Im}[\chi^{\alpha}_{ \gamma}(\omega)]&=\left[\frac{1+\alpha}{2}+n_B(\omega)\right]\rho_{ \gamma}(\omega),
\label{eqn33T}
\end{eqnarray}
where $\rho_{\gamma}(\omega)=-\frac{1}{\pi}\mathrm{Im}[\chi^R_\gamma(\omega)]$.
From here we express the bosonic correlation function, $D_{\gamma}=\sum_{\alpha}D_{\gamma}^{\alpha}$. 
The bosonic spectral functions, denoted as $\rho^\alpha_{D_{\gamma}}=-{\rm Im}(D^\alpha_\gamma)/\pi$, satisfy the following relationships:
\begin{eqnarray}
\rho_{D_{N/S}}(\omega)&=\sum_{\alpha}\rho^{\alpha}_{D_{ N/S}}(\omega); ~~~~~~~~~~~~
\rho_{D_{ N/S}}(\omega)&=-\rho_{D_{ N/S}}(-\omega); ~~~~~~~~~~~~\rho^{-}_{D_{ N/S}}(\omega)=-e^{-\beta \omega}\rho^{+}_{D_{ N/S}}(\omega).
\label{eqn28T} 
\end{eqnarray}
In addition, $\rho^{\alpha}_{D_{N}}(\omega)$ satisfies the following sum rule:
\begin{eqnarray}
\int_{-\infty}^{\infty}d\omega\rho^{\alpha}_{D_{N}}(\omega)=\alpha n.
\label{sumrule}
\end{eqnarray}
With the knowledge of the above correlation functions/spectral functions, we can rewrite the imaginary part of the two self-energies as
\begin{eqnarray}
\mathrm{Im}\left[\Sigma_{1}^{R}(\omega)\right]&=&-6\pi t^2(1-\frac{3n}{4})\int d\epsilon_1 \Big(\rho_{D_{N}}(\epsilon_1)+2\rho_{D_{S}}(\epsilon_1)\Big)\rho_{G}(\omega+\epsilon_1)\left[n_B(\epsilon_1)+n_F(\omega+\epsilon_1)\right],\\
\mathrm{Im}\left[\Sigma_{2}^{R}(\omega)\right]&=&2\pi\beta t^2\int d\epsilon_1 d\epsilon_2 \rho_{D_{N}}(\epsilon_1)\rho_{D_{S}}(\epsilon_2)\rho_{G}(\omega+\epsilon_1+\epsilon_2)\left[n_B(\epsilon_2)+n_F(\omega+\epsilon_1+\epsilon_2)\right]\left[n_B(\epsilon_1)+n_{F}(\omega+\epsilon_1)\right].
\end{eqnarray}
Finally, the single-particle fermionic Green function Eq.(\ref{eq:g_loc}) can be computed using the above two self-energies. For further details of the above in terms of $X-$operators, one can see \cite{hasan_24}.\\
\end{document}